\documentclass[prl,twocolumn,superscriptaddress,showpacs,floatfix]{revtex4}

\usepackage{graphicx}

\begin{document}               

\def\be{\begin{equation}}
\def\ee{\end{equation}}
\def\ba{\begin{eqnarray}}
\def\ea{\end{eqnarray}}
\def\bas{\begin{eqnarray*}}
\def\eas{\end{eqnarray*}}

%\addtolength{\topmargin}{2cm}

\title{Distribution of spectral widths and preponderance of spin-0
  ground states in nuclei}
\author{T.~Papenbrock}
\affiliation{Physics Division,
Oak Ridge National Laboratory, Oak Ridge, TN 37831, USA}
\affiliation{Department of Physics and Astronomy, University of Tennessee,
Knoxville, TN~37996, USA}
\author{H.~A.~Weidenm\"uller}
\affiliation{Max-Planck Institut f\"ur Kernphysik,
D-69029 Heidelberg, Germany}
\date{\today}

\begin{abstract}
We use a single j-shell model with random two-body interactions to
derive closed expressions for the distribution of and the correlations
between spectral widths of different spins. This task is facilitated
by introducing two-body operators whose squared spectral widths sum
up to the squared spectral width of the random Hamiltonian. The spin-0
width is characterized by a relatively large average value and small
fluctuations while the width of maximum spin has the largest average
and the largest fluctuations. The approximate proportionality between
widths and spectral radii explains the preponderance of spin-0 ground
states.
\end{abstract}

\maketitle

{\it Introduction:} Many regular features in the low-lying parts of
nuclear spectra can be attributed to a short-range, attractive
effective interaction with pairing and quadruple components. It thus
came as a surprise when Johnson, Bertsch and Dean~\cite{JBD98} found
that for even-even nuclei, an ensemble of nuclear shell-model
Hamiltonians with random two-body interactions is likely to yield a
spin-zero ground state. This is especially so since the probability
for a spin-zero ground state was found to be much larger than the
fraction of spin-zero states in the model space. Subsequent work
showed that similar regularities exist in bosonic~\cite{BF00} and
electronic~\cite{Jac00} many-body systems with random two-body
interactions. Thus the phenomenon of spin-0 preponderance seems a
very robust and rather generic feature.

The phenomenon has received intense attention, with reviews in
Refs.~\cite{ZVRev,ZAYRev}. Quantitative explanations for the spin-0
ground state dominance were given for exactly solvable
boson~\cite{Kus} and fermion~\cite{ZA01,Chau} systems, while
mean-field theory provides an understanding for the interacting boson
model~\cite{BFmf}. For more complex fermion systems, the situation
seems more difficult.  Approaches based on the ensemble-averaged
spectral widths, while useful for certain
systems~\cite{BFP99,KPJ00,Velazquez,PKB02,Kota}, failed to provide an
explanation for a single $j$-shell~\cite{ZVRev,ZAYRev,ZA01,YAZ02}.  Recently,
Mulhall {\it et al.}~\cite{MVZ00} found that the spectral centroids
form a band or an inverted band, depending on the sign of the
effective moment of inertia. However, it is not clear how to relate
this result to the spin of the ground state. In another approach, Zhao
{\it et al.}~\cite{ZAY01} made quantitative predictions based on the
diagonalization of the individual two-body operators.  However, the
success of this approach is not well understood.

\begin{figure}[h]
%\vskip 0.3cm
\includegraphics[width=0.45\textwidth]{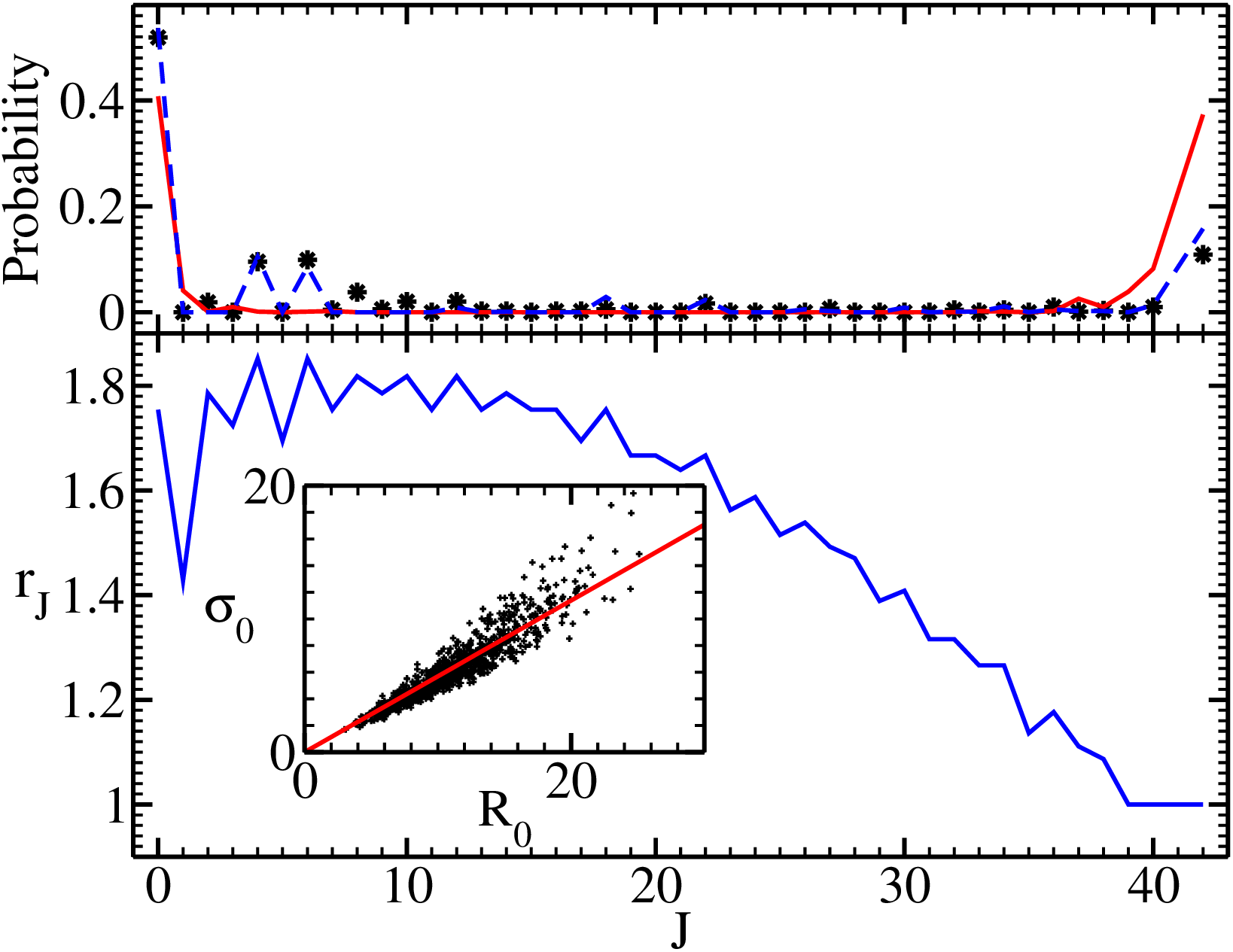}
\caption{\label{fig1}(Color online) 6 fermions in a shell with spin $j
  = 19/2$. Top: Probability that the ground state has spin $J$
  (data points); probability that spin $J$ has the largest spectral width
  (solid line); probability that the product $r_J \sigma_J$ is
  maximal (dashed line). Bottom: Scaling factor $r_J$ between the
  widths and spectral radii. Inset: Spectral radius $R_0$ versus width
  $\sigma_0$ (data points) and the linear fit (line) for total spin
  $J = 0$. (Results from 900 random realizations).}
\end{figure}

As a key to the understanding of the preponderance of spin-0 ground
states, we propose in this Letter the distribution of and the
correlations between the widths of the densities of states with
spin $J$ (in short: the spectral widths). As a model system, we
consider $n = 6$ fermions in a shell with spin $j = 19/2$ which
interact via a random two-body interaction. The data points in the
top part of Fig.~\ref{fig1} show the probability that the ground state
of this system has spin $J$. This probability is rather large for
minimum and maximum spin while the relative dimensions of the
corresponding Hilbert spaces are very small. Thus, the figure exhibits
the puzzle pointed out in Ref.~\cite{JBD98}. The solid line in the
upper part of Fig.~\ref{fig1} shows the probability that spin $J$ has
the largest spectral width. This probability is sizable only for
minimum and maximum spin. Thus, the preponderance of spin-0 ground
states is suddenly not a surprise any more! This observation prompts
us to study the distribution functions for the spectral widths, and
the correlations between spectral widths of different spins. This will
eventually lead us to a semi-quantitative understanding of the
preponderance of spin-0 ground states.

{\it The two-body random ensemble} (TBRE). We consider the
Hamiltonian matrix $H(J)$ in the $d_J$-dimensional Hilbert space
${\cal H}(J)$ of $n$-fermion states with spin $J$ for a given
two-body interaction. The latter has reduced matrix elements (TBME)
$v_\alpha, \alpha = 1, \ldots, a$. In our model system, there are $a =
10$ TBME since two fermions can have spins $j_\alpha = 0, 2, 4,
\ldots, 2j-1$. The matrix $H(J)$ is linear in the TBME,
\be
\label{hamC}
H(J)= \sum_\alpha^a v_\alpha C_\alpha(J) \ . 
\ee
The matrices $C_\alpha(J)$ transport the two-body interaction into
the Hilbert space ${\cal H}(J)$. These matrices are determined
entirely by the geometry of the shell model. They are built of
coefficients of fractional parentage and angular-momentum coupling 
coefficients.

In order to obtain generic results, we assume that the TBME are
uncorrelated Gaussian-distributed random variables with zero mean and
unit variance, $\overline{v_\alpha} = 0$ and $\overline{v_\alpha
v_\beta} = \delta_{\alpha \beta}$ where the bar denotes the ensemble
average. Our results then apply to almost all two-body interactions,
the integration measure being the product of the differentials of all
the $v_\alpha$'s. The matrices $H(J)$ are sums of random variables
and, thus, form a random-matrix ensemble, the two-body random
ensemble. For spins $J \neq J'$, the matrices $H(J)$ and $H(J')$
depend on the same random variables $v_\alpha$ and are thus
correlated.

{\it Distribution of spectral widths.} The spectral width $\sigma_J$
of the states with spin $J$, a random variable, is defined as 
\be
\label{wdef}
\sigma_J^2 = d_J^{-1} \ {\rm Tr} ( H^2(J) ) = \sum_{\alpha \beta}
v_\alpha S_{\alpha \beta} v_\beta \ .
\ee
Here $S_{\alpha\beta}(J) = d_J^{-1} {\rm Tr} (C_\alpha(J) C_\beta(J))$
is the symmetric and positive semi-definite $a$-dimensional overlap
matrix in the space of the random variables $v_\alpha$.
Diagonalization of $S = U s^2 U^T$ yields the eigenvalues $s_1^2(J) 
\ge s_2^2(J) \ge \ldots \ge s_a^2(J) \ge 0$ and the orthogonal,
$J$-dependent matrix $U$. The matrices-in-Hilbert-space
$B_\alpha(J) = \sum_\beta U_{\beta\alpha} C_\beta(J)$ are orthogonal
to each other in the sense of the trace, i.e., $d_J^{-1} {\rm
  Tr}(B_\alpha(J) B_\beta(J)) = \delta_{\alpha \beta} s^2_\alpha(J)$,
and $B_\alpha$ has the spectral width $s_\alpha$. In terms of the
$B$s, the Hamiltonian reads 
\be
\label{hamB}
H(J)= \sum_\alpha w_\alpha B_\alpha(J) \ .
\ee
The new Gaussian random variables $w_\alpha = \sum_\beta v_\beta
U_{\beta\alpha}$ obey $\overline{w_\alpha} = 0$ and
$\overline{w_\alpha w_\beta} = \delta_{\alpha \beta}$. For fixed $J$,
the random-matrix model~(\ref{hamB}) is equivalent to the
random-matrix model~(\ref{hamC}).

\begin{figure}[t]
%\vskip 0.3cm
\includegraphics[width=0.45\textwidth]{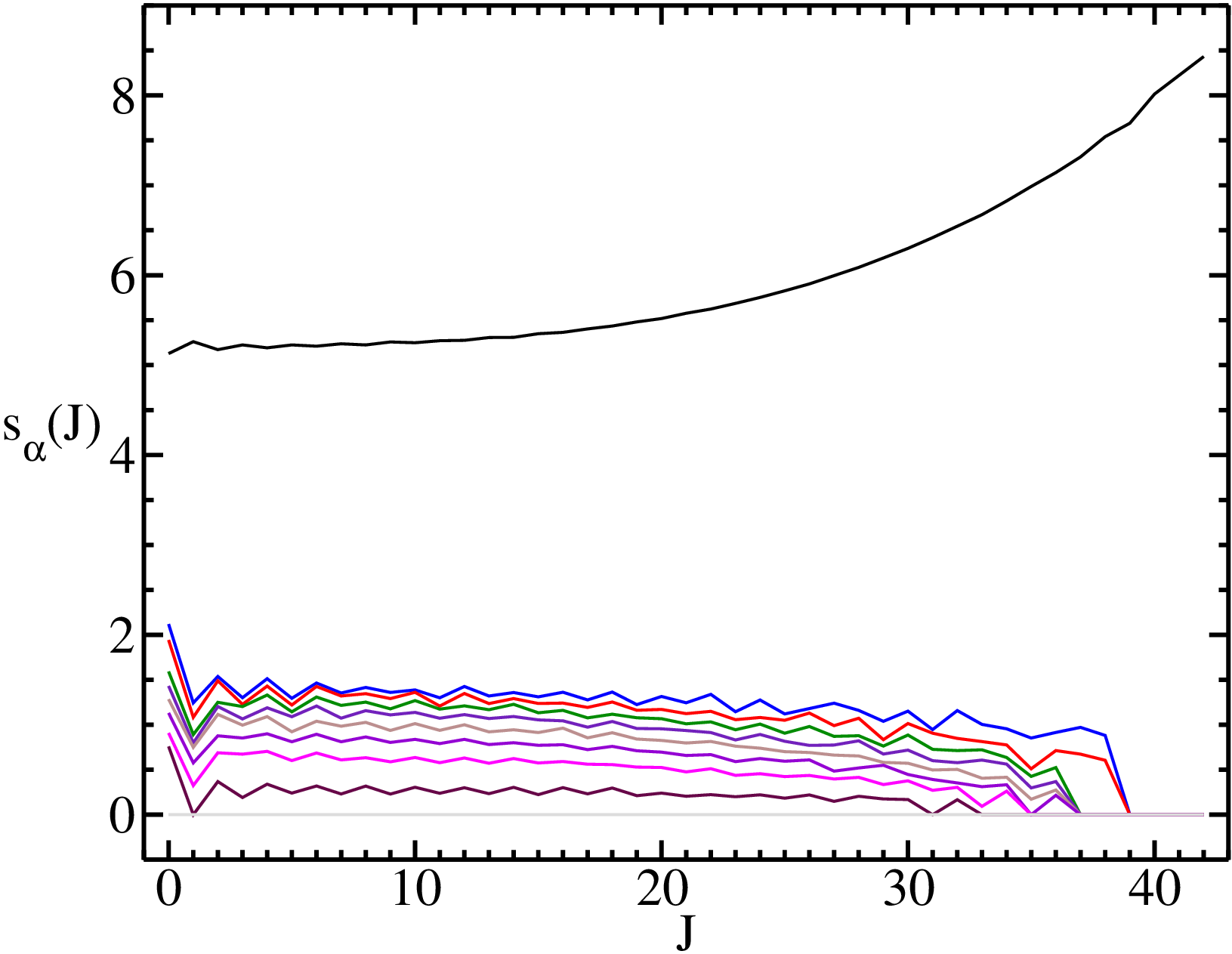}
\caption{\label{fig2}(Color online) Roots of the ten eigenvalues of
  the eigenvectors $B_\alpha(J)$ versus $J$ for 6 fermions in a $j =
  19/2$ shell.}
\end{figure}

In the new basis, we have
\be 
\label{sig2}
\sigma_J^2 = \sum_\alpha w_\alpha^2 s^2_\alpha(J).  
\ee 
The geometric aspects of the shell model are contained in the roots
$s_\alpha \geq 0$ of the eigenvalues. Figure~\ref{fig2} shows how the
$s_\alpha$ vary with spin $J$ for our model system. For each spin
$J$ there is one particularly large root $s_1$. This root increases
almost monotonically with $J$. For each spin there is one basis vector
$B_a(J)$ with root zero. This vector is given by the matrix
representation of $\hat{J}^2 - J(J+1)$ where $\hat{J}$ denotes the
spin operator. Indeed, $\hat{J}^2 - J(J+1)$ annihilates states with
spin $J$ and is a scalar two-body operator. Hence, the matrix
representation of $\hat{J}^2 - J(J+1)$ is a linear combination of the
matrices $C_\alpha(J)$ and, at the same time, an eigenvector of
$S_{\alpha \beta}$ with eigenvalue zero. For large values of $J$,
there is more than one zero eigenvalue. This is because here the
number of independent matrix elements is smaller than the number of
TBME.

We first use Fig.~\ref{fig2} for a qualitative discussion of the
distribution of widths in the TBRE. Clearly, $\overline{\sigma_J^2} =
\sum_\alpha s^2_\alpha$ attains its maximum value for maximum spin $J
= J_{\rm max}$. Among the low spins, $\overline{\sigma_0^2}$
dominates because the root $s_1$ is almost constant for low spins, and
the remaining non-zero roots are exceptionally large for $J = 0$.
As for the fluctuations, we use two limiting cases for orientation:
(i) All roots are equal to $s^2$. Then $\overline{\sigma^2} = a s^2$
and the r.m.s. variance is $\sqrt{2 a} s^2$. (ii) Only the root
$s_1$ differs from zero. Then $\overline{\sigma^2} = s^2_1$ and the
r.m.s. variance is $\sqrt{2} s^2_1$. Thus, we expect that in relation
to the average width, the r.m.s. fluctuations of $\sigma_0$ are
smaller than those of the $\sigma_J$ for large $J$.

\begin{figure}[t]
%\vskip 0.3cm
\includegraphics[width=0.45\textwidth]{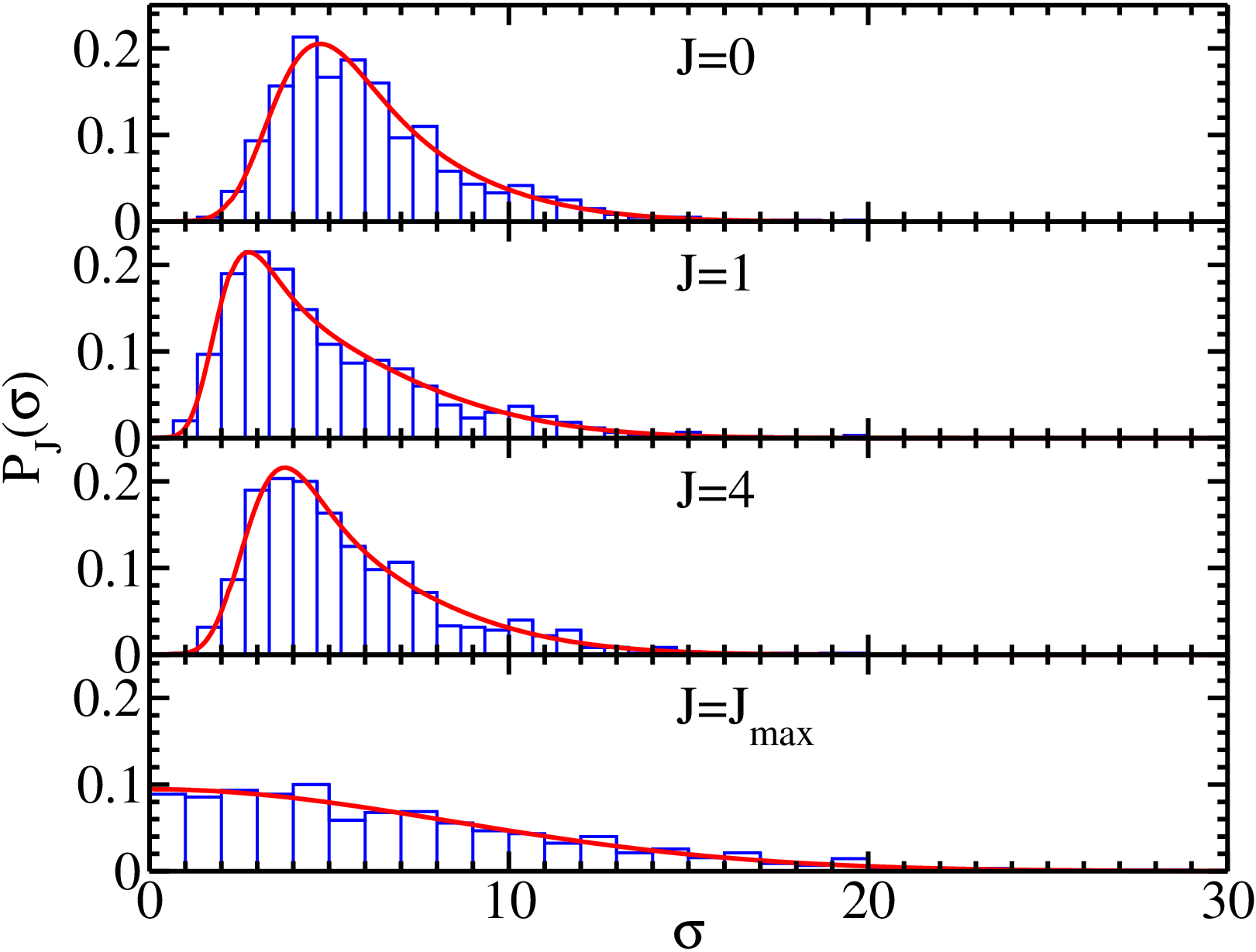}
\caption{\label{fig3}(Color online) Normalized probability
  distribution functions $P_J(\sigma)$ versus spectral width $\sigma$.
  The histograms result from 900 random realizations for 6 fermions in
  a $j = 19/2$ shell. The line shows the theoretical
  result~(\ref{final}).}
\end{figure}

For a quantitative analysis, we define the probability distribution
$P_J(\sigma) = 2 \sigma \overline{\delta( \sigma^2-\sigma_J^2 )}$ for
finding a value of the width between $\sigma$ and $\sigma+{\rm
  d}\sigma$. The factor $2\sigma$ stems from the differential ${\rm d}
\sigma^2 = 2 \sigma {\rm d} \sigma$. For $J = J_{\rm max}$ there is
only one non-zero root $s_1$, and $P_{J_{\rm max}}$ is a Gaussian
with width $s_1$. For general $J$, all the integrations over the
$w_\alpha$'s can be done, and the expression for $P_J(\sigma)$
reduces to
\be
\label{final}
P_J(\sigma)={\sigma\over\pi}
\int\limits_{-\infty}^\infty {\rm d} t e^{i t \sigma^2}
\prod_{\alpha = 1}^a \frac{e^{-{i \over 2} \arctan{ 2 t s^2_\alpha}}}
{\left(1 + 4s_\alpha^4 t^2 \right)^{1/4}} \ .
\ee
Figure~\ref{fig3} shows $P_J(\sigma)$ for several spins $J$. This
agrees very well with the histograms from 900 realizations of our
TBRE.

{\it  Correlations of widths.} The correlations between the
Hamiltonian matrices $H(J)$ and $H(J')$ induce correlations between
the widths $\sigma_J$ and $\sigma_{J'}$. This fact is borne out by the
numerical simulation: Under the assumption that width correlations can
be neglected, we can use the distributions $P_J(\sigma)$ to compute
the probability that spin $J$ has a larger width than spin $J'$. Such
a calculation yields, e.g., that $\sigma_0 > \sigma_4$ with 60\%
probability. The numerical simulation of the TBRE shows, however, that
$\sigma_0 > \sigma_4$ for 93\% of all realizations. Hence, there
must be strong correlations between the widths.

\begin{figure}[t]
%\vskip 0.3cm
\includegraphics[width=0.45\textwidth]{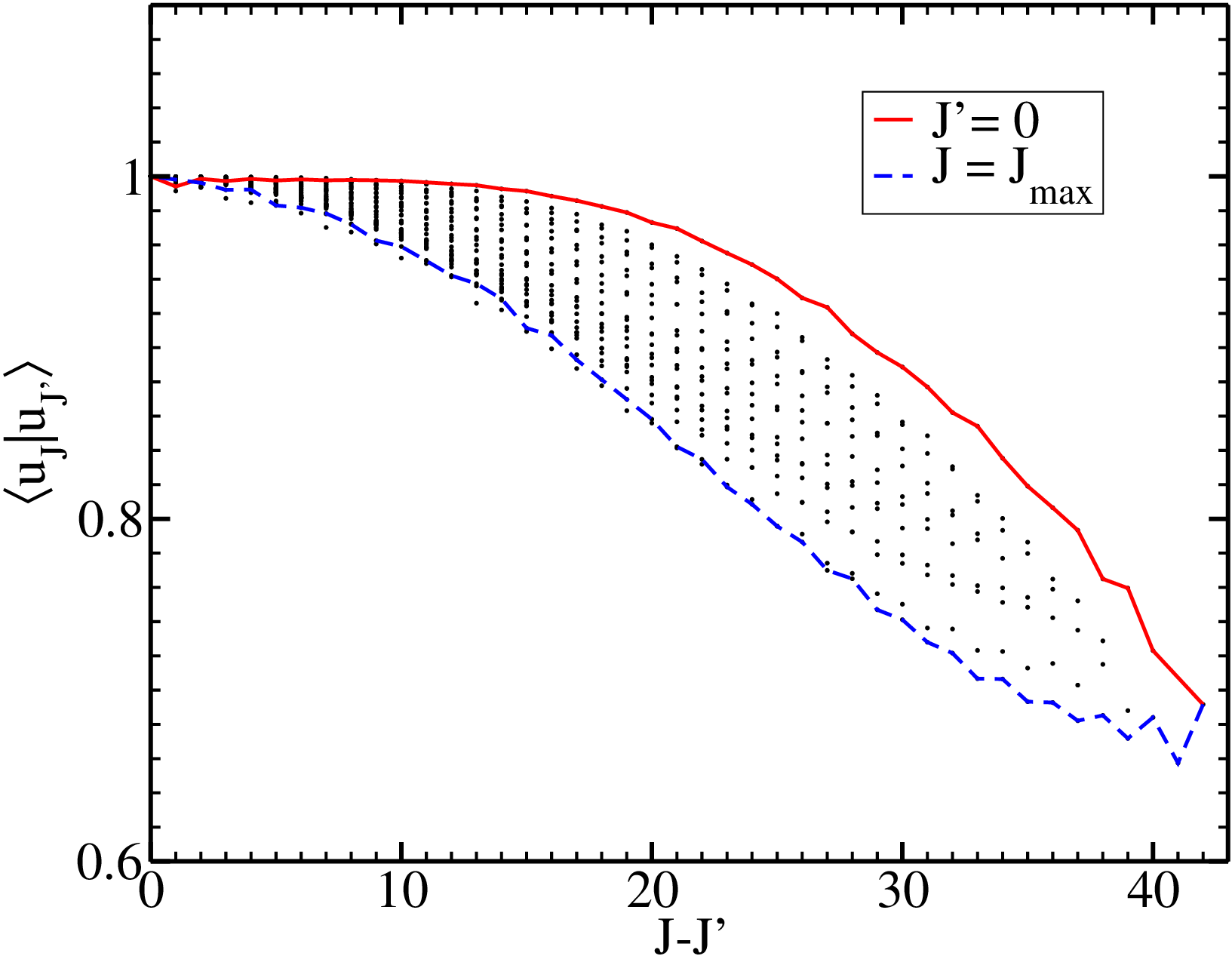}
\caption{\label{fig4}(Color online) Data points: Values of the scalar
  products $\langle u_J | u_{J'} \rangle$ of the largest-width
  eigenvectors $|u_J \rangle$ of the overlap matrices $S_{\alpha
  \beta} = d_J^{-1}{\rm Tr}(C_\alpha(J) C_\beta(J))$ for $J \ge J'$.
  The upper and lower line connect the data points corresponding to
  $J' = 0$ and $J = J_{\rm max}$, respectively.}
\end{figure}

We again begin with a qualitative analysis. Let $|u_J \rangle$ denote
that eigenvector of the overlap matrix $S_{\alpha \beta}(J)$ which
corresponds to the largest root $s_1(J)$. Figure~\ref{fig4} shows the
scalar products $\langle u_J | u_{J'} \rangle$ as functions of $J -
J'$. Clearly, the eigenvectors $|u_J \rangle$ depend only very weakly
on $J$ and are almost identical for total spins that differ by just a
few units. In particular, $\langle u_J | u_{J'} \rangle$ decreases
very slowly with $J$ for $J' = 0$ (upper line in Fig.~\ref{fig4})
while the decrease is faster for $J = J_{\rm max}$ (lower line in
Fig.~\ref{fig4}). We see that for low spins up to values of 10 to 20,
the eigenvector of the largest root and the largest root itself are
almost independent of $J$. Because of these strong correlations among
roots of low spins, the contributions of the largest roots $s_1$ to
the widths are simultaneously small or simultaneously large for most
realizations of the TBRE. In this way the correlations significantly
enhance the probability that spin $J = 0$ has the largest total width
among all low spins. Similar arguments show that the maximum spin $J =
J_{\rm max}$ is most likely to exhibit the largest width for high
spins. These statements are in keeping with the numerical results
presented in the upper part of Fig.~\ref{fig1}.

For quantitative results, we define the probability $p(J,J') =
\overline{\Theta \left( \sigma_J^2 - \sigma_{J'}^2 \right)}$ that
spin $J$ has a larger width than spin $J'$, where $\Theta$ denotes the
unit step function. Introducing the eigenvalues $q_\alpha$ of the
matrix $S_{\alpha \beta}(J) - S_{\alpha \beta}(J')$, we can perform
the ensemble average. Employing integral transforms, we arrive at
\be
\label{w>}
p(J,J')={1\over 2} + 
{1\over \pi} \int\limits_{0}^\infty {\rm d}t 
{\sin{\left( {1 \over 2} \sum_{\beta} \arctan{2 t q_\beta}\right)}
\over 
t \prod_{\alpha} \left(1+4q_\alpha^2 t^2\right)^{1/4}} \ .
\ee
This expression accurately describes the results of our
numerical simulation. 

{\it From widths to ground states.} The bridge between the spectral
widths discussed so far and the spin of the ground state is spanned by
the scaling factors $r_J$. These are defined as follows. For each
value of $J$, every realization of the TBRE yields a spectrum
$E_{J,i}, i=1, \ldots, d_J$. Let $R_J = \max_{i}|E_{J,i}|$ be the
spectral radius for spin $J$. For each realization, $R_J$ is given
with equal probability either by the energy of the lowest-lying level
or by that of the highest-lying level with spin $J$. The spectral
radius is related to the spectral width $\sigma_J$ of that realization
by the inequality $R_J \ge \sigma_J$. However, our numerical results
show that the much stronger relation \be
\label{R2Tr}
R_J \approx r_J \ \sigma_J
\ee
is approximately valid, with $r_J$ constant (independent of the
realization considered). This is shown in the inset of
Fig.~\ref{fig1}. The scaling factor $r_J \ge 1$ (shown in the lower
part of Fig.~\ref{fig1}) encodes information contained in the tails of
the spectral density. The average level density typically decreases
exponentially fast for energies close to the spectral radius. Thus,
$r_J$ is expected to exhibit only a logarithmically weak dependence on
the dimension $d_J$ and on the average width. We neglect this
dependence as well as the difference between the absolute values of
the energies of the highest and the lowest level in the spectrum of
each realization. Thus, we identify the product $r_J \sigma_J$ with
the energy of the lowest-lying state with spin $J$. It is then
tempting to assume that the probability distribution of $R_J$ is
simply given by that of $\sigma_J$ scaled by the factor $r_J$. Our
numerical results show, however, that the two distributions, although
similar, do not really coincide. However, the relation~(\ref{R2Tr}) is
sufficiently accurate to determine reasonably reliably the probability
that $r_J \sigma_J$ is maximal. Indeed, this probability is plotted
as the dashed line in the upper part of Fig.~\ref{fig1}. The agreement
with the data points is very satisfactory.

We have obtained similar results for $n = 8$ fermions in a $j = 19/2$
shell. For the odd-number system with $n = 7$, we also found that
minimum and maximum spins are most likely to exhibit the largest
widths. However, other spins also have smaller but sizable
probabilities of having the largest widths.

{\it Summary}. We have studied the spin of the ground state for the
shell model with random two-body interactions. To this end, we have
derived closed expressions for both, the width distribution functions
and the width correlation functions. Using these results, we have
shown that spin-0 and maximum spin are most likely to exhibit the
largest widths. The spin-0 width is characterized by a relatively
large average value and rather small fluctuations, while the maximum
spin displays the largest average and the largest fluctuations. We
have numerically established an approximate proportionality between
spectral widths and spectral radii. This relation is sufficiently
reliable to explain the preponderance of spin-0 ground states in the
shell model.

This work has been confined to a single shell with spin $j = 19/2$.
However, our results depend only upon geometrical properties of that
model and are thus expected to apply similarly in other shells. We
speculate that similar considerations would also apply to other
many-body systems.

The distribution functions were calculated using superpositions of
two-body operators whose squared widths sum up to the average
spectral width of the random Hamiltonian. These operators may have
further applications in nuclear structure calculations: Effective
interactions are often obtained from fitting two-body matrix
elements~\cite{Honma02}. These procedures ideally should avoid
generating linear combinations of two-body operators with small or
zero spectral widths.

Our work poses the intriguing analytical problem to derive the
eigenvalues $s_\alpha(J)$. These are determined entirely by the
geometry of the shell model. Hence, the values of the $s_\alpha(J)$
should be accessible by group-theoretical techniques. An
understanding of the scaling factors $r_J$ seems less difficult. It
would require the knowledge of the shapes of the average spectra.
This would allow us to determine $r_J$ from the position of the lowest
level (found by integrating the normalized spectrum from $- \infty$ to
where the integral equals $1/2$) and the width of that spectrum. 

This research was supported in part by the U.S. Department of Energy
under Contract Nos.\ DE-FG02-96ER40963 (University of Tennessee) and
DE-AC05-00OR22725 with UT-Battelle, LLC (Oak Ridge National
Laboratory).

\end{document}